\RequirePackage{fix-cm}
\documentclass[smallextended]{svjour3}       % onecolumn (second format)
\smartqed  % flush right qed marks, e.g. at end of proof
\usepackage[all]{xy}
\usepackage{graphicx}
\usepackage{amssymb}            
\usepackage{amsmath}
\usepackage{todonotes}
\usepackage{float} 

\usepackage[latin1]{inputenc}

\newcommand{\OO}{\mathrm{O}}                % Orthogonal group

\newcommand{\V}{\underline}

\newcommand{\bee}{\begin{equation}}
\newcommand{\eee}{\end{equation}}
\newcommand{\ben}{\begin{equation*}}
\newcommand{\een}{\end{equation*}}
\newcommand{\ban}{\begin{eqnarray*}}
\newcommand{\ean}{\end{eqnarray*}}
\journalname{}
\begin{document}

\title{On the isotropic moduli of 2D strain-gradient elasticity%\thanks{Grants or other notes
}


\author{N. Auffray}


\institute{ N. Auffray \at
            MSME, Universit\'{e} Paris-Est, Laboratoire Mod\'{e}lisation et Simulation Multi Echelle,\\
            MSME UMR 8208 CNRS, 5 bd Descartes, 77454 Marne-la-Vall\'{e}e, France\\
              \email{Nicolas.auffray@univ-mlv.fr} }

\date{\today}

\maketitle

\begin{abstract}
In the present paper, the simplest model of  strain-gradient elasticity will be considered, that is the isotropy in a bidimensional space. Paralleling the definition of the classic elastic moduli, our aim is to introduce second-order isotropic moduli having a mechanical interpretation. A general construction process of these moduli will be proposed.
As a result it appears that many sets can be defined, each of them constituted of 4 moduli: 3 associated with 2 distinct mechanisms and the last one  coupling these mechanisms. We hope that these moduli (and the construction process) might be useful for the forthcoming investigations on strain-gradient elasticity.

\keywords{Group representation \and harmonic decomposition \and strain-gradient elasticity}
\end{abstract}

\section{Intro}

Until recently and, for example, the growing industrial interest in composite structures, materials studied in engineering science were paradigmatically considered isotropic and  homogeneous. For industrial needs, the modeling effort was not focused on a fine comprehension of the elastic behavior but rather on the description of overall non-linear phenomena, such as: plasticity, damaging, fatigue...
These last years, the point of view has changed. And if the modeling interests remain focused on the same topics, the needs for new innovative materials lead scientists to reconsider the architecture of the matter \cite{BBE08,FDA10}. By architecture we deliberately choose not to focus on a specific scale of  organization of the matter. This organization may as well concern the microstructure of an alloy \cite{BSP09},  as commonly conceived, or any other intermediate scale. For structural applications, a well-known example of mesoscale organized material is the honeycomb structure \cite{GAS+82}. These last years, in engineering science, the classical boundary between material and structure has been blurred  \cite{MDP+12,BE12}, and  the center of interest has been shifted from homogeneity to heterogeneity and from isotropy to anisotropy.  Such an evolution has been sustained by important progress make both in the fabrication of material (e.g. rapid prototyping, or spark plasma sintering \cite{FMB+12}),  and experimental testing devices (digital image correlation \cite{RRH09}, infrared correlation \cite{MPL+12}, $\ldots$). To some extent, engineering scientists model materials which can be elaborated.

But what is a paradigm shift in engineering science, is the ground state in some other mechanical fields. In  biomechanics, in order to spare matter, materials are almost always multifunctional \cite{FW07}. This multifunctionality is a consequence of the organization of the matter at different scales. In bone cortical structure at least 5 different scales can been identified \cite{SNB+10} and in the butterfly wing,  the multilayered structure results both in pigment-free coloration \cite{VVB+12} and super hydrophobic properties \cite{ZGJ07}. Needless to say that biomaterials are neither homogeneous nor isotropic.
Another field in which heterogeneity and anisotropy are the rules is obviously internal geophysics \cite{Hel96b}. Be it for geological prospecting \cite{BG68,BG70}, for seismic tomography \cite{Hel96} or other needs, geophysicians has, since the origin, studied anisotropic elasticity, both in infinitesimal and finite deformations \cite{Boe77,Boe79}.
As a consequence, the actual corpus of knowledge in theoretical elasticity has been deeply developed by these two communities \cite{Bac70,BBS07,Boe79,Cow89,MC90}. 

In the present paper attention is devoted to generalized elasticity, more precisely to Strain Gradient Elasticity (SGE in the following) as introduced by Mindlin \cite{Min64,Min65,ME68}. This theory is a way, among others, to incorporate effects of characteristic lengths of materials microstructure into constitutive models.. 
Up to recently, and due to their intrinsic difficulties (both theoretical and computational) generalized continuum models were mostly confined to theoretical discussions. Recently, a regain of interest toward these theories has been observed. Their applications have been investigated for the description of bones \cite{YL82,FVO02,MGL+12}, of granular materials \cite{Ahm82,MV87}, to describe localization in geomaterials \cite{CCM01}, to study the consolidation of soil layers \cite{GII+08,MII+08}, and so on.

Such a rebirth has been sustained both by an increasing interest in micro-macro models and the augmentation of the computational power. But even tough these models arouse new interest, their mechanical contents remain unclear. The objective of the present contribution is to give a better insight into the physics contained in SGE. To that aim, attention will be restricted  the simplest  case of bidimensionnal isotropy. As well-known, in the same situation classical elasticity is characterized by a couple of parameters, which can either be 
\begin{itemize}
\item $(E,\nu)$: the Young modulus and the Poisson ratio;
\item $(\lambda,\mu)$: the Lamé coefficients;
\item $(K,\mu)$: the bulk and shear modulus.
\end{itemize}
and so on since there is an infinite number of possibilities. For practical applications, the use of the moduli has to be preferred to the use of tensor components (for example, and among others, the  couple $(c_{1111}, c_{1122})$). The reason is that, contrary to tensor entries, isotropic moduli are (most of the time) associated with a mechanical content.  Concerning SGE,  and even if their matrix representations are known both in 2D and 3D spaces \cite{ABB09,ISV09,ALH13}, analogues to isotropic moduli have not been clearly identified. The aim of the present paper is to fill this gap. As it will be seen, SGE can be (non-uniquely) decomposed into 2 distinct physical mechanisms characterized by 4 moduli: one of the mechanisms is defined by 2 moduli, the other is defined by 1 modulus, while the last modulus couples these mechanisms. Similarly to classical elasticity this set of  moduli is not uniquely defined.\\

\noindent The paper is organized as follows. In the first section, the basic equations of SGE are provided together with a matrix representation of the isotropic constitutive law. In \S.\ref{eq:HarDec} the algebraical structure of the space of strain-gradient and hyperstress tensors is investigated. To that aim some tools stemming from group representation theory, and introduced in mechanics by Backus \cite{Bac70}, are used. As a consequence a generalization of the deviatoric and spheric decomposition for strain-gradient tensors tensors is obtained.
This decomposition is not unique and, among all the possibilities, we propose two different physically-based constructions for it. Using these results, in \S.\ref{s.MatSys}, the matrix representation for the isotropic SGE is expressed in the two different bases previously determined. In these new systems the matrices are almost diagonal, and the associated moduli have now a mechanical interpretation. Among the 4 moduli, 3 of them are associated with 2  mechanisms while the last one is a coupling parameter between these mechanisms. Finally, these new matrices will be compared with the diagonalization of the original isotropic operator, showing that one of these moduli is also an eigenvalue of the operator.\\

Let us define some notations that will be used throughout the paper. Some other less important notation  will be directly introduced in the core of the text. The following matrix groups will be considered in the present paper:
\begin{itemize}
\item $\mathrm{GL}(2)$: the group of all the invertible transformations of $\mathbb{R}^{2}$, i.e. if $\mathrm{F}\in\mathrm{GL}(2)$ then $\mathrm{det}(\mathrm{F})\neq0$;
\item $\mathrm{O}(2)$: the orthogonal group, i.e. the group of all isometries of $\mathbb{R}^{2}$ i.e. if $\mathrm{Q}\in\mathrm{O}(2)$ $\mathrm{det}(\mathrm{Q})=\pm 1$ and $\mathrm{Q}^{-1}=\mathrm{Q}^{T}$, where the superscript $^{T}$ denotes the transposition;
\item $\mathrm{SO}(2)$: the special orthogonal group, i.e. the subgroup of $\mathrm{O}(2)$ of elements satisfying $\mathrm{det}(\mathrm{Q})= 1$. This is the group of 2D rotations.
\end{itemize}

Vector spaces will be denoted using blackboard bold fonts, and their tensorial order indicated by using formal indices. Generic tensor spaces will be denoted $\mathbb{T}$, and $k$-th order  harmonic tensor spaces: $\mathbb{H}^{k}$. Harmonic tensor spaces are $\mathrm{O}(2)$-invariant, their elements are \emph{completely symmetric} and \emph{traceless} $k$-order tensors \cite{Bac70,JCB78,Bae93,FV96}. In 2D we have for $k>0$ $\mathrm{dim}(\mathbb{H}^{k})=2$, and $\mathrm{dim}(\mathbb{H}^{0})=1$. The precise definitions of these spaces will be detailed in the text.

When needed, index symmetries are expressed as follows:  $(..)$  indicates invariance under permutation of the indices in parentheses, and $\underline{..}$ indicates invariance with respect to permutations of the underlined blocks. In the following, classically the spatial derivation will be denoted by $\nabla$.

\section{The isotropic strain-gradient elasticity}\label{s:eqSGE}

\subsection{Constitutive law}

In the strain-gradient theory of linear elasticity (see, e.g., \cite{Min64,ME68}), the constitutive law gives the symmetric Cauchy stress tensor $\mathbf{\sigma }$ and the hyperstress tensor $\mathbf{\tau }$ in terms of the infinitesimal strain tensor $\mathbf{\varepsilon }$\ and strain-gradient tensor $\mathbf{\eta}=\nabla \mathbf{\varepsilon }$ through the two linear relations:
\begin{equation}
\begin{cases}
\sigma _{ij}=C_{ijlm}\varepsilon _{lm}+M_{ijlmn}\eta_{lmn}, \\ 
\tau _{ijk}=M_{lmijk}\varepsilon _{lm}+A_{ijklmn}\eta_{lmn}.%
\end{cases}
\label{SGE}
\end{equation}%
in which $\sigma _{ij}$, $\varepsilon _{ij}$, $\tau _{ijk}$ and $\eta_{ijk}=\varepsilon _{ij,k}$ denotes, respectively, the matrix components of $\mathbf{\sigma }$, $\mathbf{\varepsilon}$, $\mathbf{\tau }$ and $\mathbf{\eta }$ relative to an orthonormal basis $\{\mathbf{e}_{1},\mathbf{e}_{2}\}$ of $\mathrm{E}^{2}$, the two-dimensional Euclidean space; $C_{ijlm}$, $M_{ijlmn}$ and $A_{ijklmn}$\ are the matrix components of the fourth-, fifth- and sixth-order elastic stiffness tensors $\mathrm{\mathbf{C}}$, $\mathrm{\mathbf{M}}$ and $\mathrm{\mathbf{A}}$, respectively. These matrix components satisfy the following index permutation symmetry properties:%
\ben	
C_{\underline{(ij)}\  \underline{(lm)}}\quad ; \quad M_{(ij)(kl)m}\quad ;\quad A_{\underline{(ij)k}\  \underline{(lm)k}}
\label{Sym1}
\een
in which the minor symmetries ($(..)$) are intrinsic symmetries of the tensors, while the major ones ($\underline{..}$) are due to the existence of an elastic potential.
In the case of a centro-symmetry material, the fifth-order elastic stiffness tensor $\mathrm{\mathbf{M}}$ vanishes, and the constitutive laws (\ref{SGE}) become uncoupled:
\begin{equation}
\begin{cases}
\sigma _{ij}=C_{ijlm}\varepsilon _{lm}, \\ 
\tau _{ijk}=A_{ijklmn}\eta_{lmn}.%
\end{cases}
\label{Uncpd}
\end{equation}
The \emph{explicit matrix forms} of $\mathrm{\mathbf{A}}$ have been determined, both in 2D and 3D, for all symmetry classes \cite{ISV09,ABB09,ABB10}. Here only the results needed for the present study will be summed up.

\subsection{Orthonormal basis and matrix component ordering}

\label{ss:OrtOrd}

Let us consider the following space:
\ben
\mathbb{T}_{(ij)k}=\{\mathrm{T}\in\mathbb{T}_{ijk}|\mathrm{T}=\displaystyle\sum\limits_{i,j,k=1}^{2}T_{ijk}\mathbf{e}_{i}\otimes \mathbf{e}_{j}\otimes\mathbf{e}_{k},\ T_{ijk}=T_{jik}\} 
\een
which is, in 2D, a 6-dimensional vector space. The SGE tensor $\mathrm{\mathbf{A}}$ is a self-adjoint endomorphism of $\mathbb{T}_{(ij)k}$.
In order to express the strain-gradient $\mathbf{\eta}$ and the hyperstress tensor $\mathbf{\tau}$ as 6-dimensional vectors and write $\mathrm{\mathbf{A}}$ as $6\times 6$ matrix, we introduce the following orthonormal\footnote{In $\mathbb{R}^{6}$, we naturally consider the standard euclidean norm, that is the norm induced by the scalar product. In other terms, the basis we construct satisfy the property $\mathbf{\widehat{e}}_{\alpha }\cdot\mathbf{\widehat{e}}_{\beta}=\delta_{\alpha\beta}$.} basis vectors:
\ben
\mathbf{\widehat{e}}_{\alpha }=\left( \frac{1-\delta _{ij}}{\sqrt{2}}+\frac{\delta _{ij}}{2}\right) \left( \mathbf{e}_{i}\otimes \mathbf{e}_{j}+\mathbf{e}_{j}\otimes \mathbf{e}_{i}\right) \otimes \mathbf{e}_{k},\quad 1\leq \alpha\leq 6  
\een
where the Einstein summation convention does not apply.
Then, the aforementioned tensors can be expressed as:
\ben
\mathbf{\eta} = \displaystyle\sum\limits_{\alpha=1}^{6}\widehat{\eta}_{\alpha}\mathbf{\widehat{e}}_{\alpha}, \quad  \mathbf{\tau} = \displaystyle\sum\limits_{\alpha=1}^{6}\widehat{\tau}_{\alpha}\mathbf{\widehat{e}}_{\alpha}, \quad 
\mathrm{\mathbf{A}}=\displaystyle\sum\limits_{\alpha,\beta=1,1}^{6,6} \widehat{A}_{\alpha\beta}\mathbf{\widehat{e}}_{\alpha}\otimes \mathbf{\widehat{e}}_{\beta },    
\een
so that the second relation in (\ref{Uncpd}) can be conveniently written in the matrix form%
\ben
\widehat{\tau}_{\alpha}=\widehat{A}_{\alpha\beta}\widehat{\eta}_{\beta}
\label{SGER}
\een%
Using the introduced orthonormal basis, the relations between the matrix components $\widehat{\eta}_{\alpha}$ and $\eta_{ijk}$ are
\ben
\widehat{\eta}_{\alpha }=%
\begin{cases}
\eta_{ijk}\text{ \ if \ }i=j, \\ 
\sqrt{2}\eta_{ijk}\text{ \ if \ }i\neq j;%
\end{cases}%
\label{3-to-1}
\een
and, obviously, the same relations between $\widehat{\tau}_{\alpha}$ and $\tau_{ijk}$ hold. For the constitutive tensor we have the following relations: 
\ben
\widehat{A}_{ \alpha\beta }=%
\begin{cases}
A_{ijklmn}\text{ \ if \ }i=j\ \text{and}\ k=l\text{,} \\ 
\sqrt{2}A_{ijklmn}\text{ \ if \ }i\neq j\ \text{and}\ l=m\ \text{or}\ i=j\ 
\text{and}\ l\neq m, \\ 
2A_{ijklmn}\text{ \ if \ }i\neq j\ \text{and}\ l\neq m.
\end{cases}
\label{6-to-3}
\een
It remains to choose an appropriate three-to-one subscript correspondence convention between $ijk$ and $\alpha$. The correspondence specified in Table 1 was chosen in order to make the 6th-order tensor $\mathrm{\mathbf{A}}$ block-diagonal for dihedral classes \cite{ABB09,ABB10}.
\begin{table}[H]
\begin{center}
\begin{tabular}{|c||c|c|c||c|}
\hline
$\alpha$& $1$& $2$& $3$&  \\ \hline\hline
$ijk $&$111 $&$221 $&$122$& Privileged direction: $1$\\ 
\hline\hline
$\alpha $&$6$&$7$&$8$&  \\ \hline
$ijk $&$222$&$112$&$121 $& Privileged direction: $2$\\ 
\hline\end{tabular}
\end{center}
\caption{The three-to-one subscript correspondence for 2D strain-gradient/hyperstress tensors}
\end{table}
In the present study attention will be focused on the isotropic class. Therefore using this representation, the following matrix is obtained:
\begin{equation}\label{Aiso}
\mathrm{\mathbf{A}}_{\OO(2)}=
\begin{pmatrix}
 a_{11}& a_{12}& \frac{a_{11}-a_{22}}{\sqrt{2}}-a_{23}& 0                       &0&0\\
      & a_{22}& a_{23}                               &0                 &0     &0\\     
      &      & \frac{a_{11}+a_{22}}{2}-a_{12}       & 0 & 0& 0\\
      &      &                                     & a_{11}& a_{12}& \frac{a_{11}-a_{22}}{\sqrt{2}}-a_{23}\\
      &      &                                     &      & a_{22}& a_{23}\\       
       &&&&&		\frac{a_{11}+a_{22}}{2}-a_{12}
\end{pmatrix}
\end{equation}
It can readily be observed that an isotropic SGE tensor is defined by a set of 4 matricial components\footnote{Obviously this choice is not unique and many other sets of components might have  been retained.}: ${a_{11},a_{12},a_{22},a_{23}}$.   Even if, the matrix is well structured, it is difficult to see the mechanical content of those isotropic components. To obtain a better insight into the physics described by this tensor, pertinent isotropic moduli have to be defined. To proceed toward this goal, the algebraical structure of $\mathbb{T}_{(ij)k}$ has to be investigated. 

\section{The harmonic decomposition of  $\mathbb{T}_{(ij)k}$}\label{s:HarDec}

In 2D, the decomposition of a tensor space into a direct sum of $\mathrm{O}(2)$-irreducible spaces is the generalization to higher-order tensors of the deviatoric and spheric decomposition of the space of symmetric second-order tensors. This decomposition will be referred to as the harmonic decomposition. $\mathbb{T}_{(ij)k}$ can be decomposed\footnote{It is worth pointing out that the structure of the harmonic decomposition depends on the space dimension. In 3D the decomposition is different and takes the following form
\ben\label{eq:HarDec}
\mathbb{T}_{(ij)k}\cong \mathbb{H}^{3}\oplus\mathbb{H}^{\sharp 2}\oplus\mathbb{H}^{1}_{a}\oplus\mathbb{H}^{1}_{b}
\een
in which $\dim\mathbb{H}^{k}=2k+1$ and where the $\sharp$ superscript indicates a space which elements change sign if the space orientation is reverse. In other terms the $\sharp$ superscript indicates spaces of pseudo-tensors.} into the direct sum of three $\mathrm{O}(2)$-irreducible spaces: 
\bee\label{eq:HarDec2}
\mathbb{T}_{(ij)k}\cong \mathbb{H}^{3}\oplus\mathbb{H}^{1}_{a}\oplus\mathbb{H}^{1}_{b}
\eee
In 2D, each harmonic space $\mathbb{H}^{k}$ is bidimensional and contains complete symmetric traceless tensors. The superscript $k$ indicates how elements of the space transform under a rotation. For example, if a rotation of angle $\theta$ is applied, elements of $\mathbb{H}^{k}$ will be rotated by an angle $k\theta$. This decomposition is analogous to a Fourier series expansion in which the spaces $\mathbb{H}^{k}$ contain $k$-th order harmonics. This decomposition is irreducible, none of the spaces can be decomposed any further.
Thus, any element of $\mathbb{T}_{(ij)k}$ can be decomposed into $3$ elementary mutually orthogonal parts. To avoid any confusion in naming the elements of the decomposition the designation\footnote{Due to its importance in mechanics, it is important to note that  $\mathbb{H}^{2}$ is the space of deviators.} used by Jerphagnon and coworkers will be used \cite{JCB78}. Thus $\mathbb{T}_{(ij)k}$  can be decomposed into:
\begin{itemize}
\item a septor, i.e  an element of the seven-dimensional space $\mathbb{H}^{3}$;
\item two vectors which belong respectively to $\mathbb{H}^{1}_{a}$ and $\mathbb{H}^{1}_{b}$.
\end{itemize}
It is worth noting that the harmonic decomposition is uniquely defined if and only if the order of all the harmonic spaces in the decomposition are different \cite{GSS88}. For $\mathbb{T}_{(ij)k}$ since two vectors are present the decomposition is not unique. The septor part is canonical but there is an ambiguity in the definition of the vector parts.
Nevertheless among the possible constructions some are more natural than others since they carry relevant information on the physics encoded by strain-gradient tensors. In the next subsection two explicit decompositions will be defined based on \emph{mechanical} considerations.  These decompositions differ by the interpretation of the vector parts. It should be emphasized that other constructions are possible, which might be more natural according to the situation.

\subsection{The two decompositions}

These two constructions are defined as follow:
\begin{itemize}
\item The first construction consists in splitting $\mathrm{T}_{(ij)k}$ into a full symmetric tensor and a remainder one before proceeding to the harmonic decomposition of each part separately.
This way to proceed has a meaning for any third-order tensor.
\item The second construction is based on the derivation of the harmonic decomposition of $\mathrm{T}_{(ij)}$, and therefore has a real meaning only if the third-order tensor is the gradient\footnote{Or, by duality, the first moment of a second order tensor.} of some second order tensor.
\end{itemize}

\subsubsection*{Third-order tensor approach}

This approach is summed-up by the following diagram:
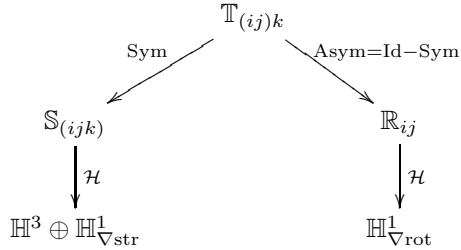
\ben
\xymatrix{
       &  \mathbb{T}_{(ij)k} \ar[ld]_{\mathrm{Sym}}\ar[rd]^{\mathrm{Asym}=\mathrm{Id}-\mathrm{Sym}} &  \\
  \mathbb{S}_{(ijk)} \ar[d]^{\mathcal{H}}  &    & \mathbb{R}_{ij} \ar[d]^{\mathcal{H}} \\
  \mathbb{H}^{3}\oplus\mathbb{H}^{1}_{\nabla \mathrm{str}}&&\mathbb{H}^{1}_{\nabla \mathrm{rot}}}
\een
where $\mathrm{Sym}$, $\mathrm{Asym}$ and $\mathcal{H}$ stands,respectively  for the symmetrization, anti-symmetrization and the harmonic decomposition process.
$\mathrm{T}_{(ij)k}$ is first split into a full symmetric tensor and a remainder one:
\ben
T_{(ij)k}=S_{ijk}+\frac{1}{3}\left(\epsilon_{ik}\epsilon_{pj}V^{\nabla \mathrm{rot}}_{p}+\epsilon_{jk}\epsilon_{pi}V^{\nabla \mathrm{rot}}_{p}\right)
\een 
where $\epsilon_{ij}$ denotes the Levi-Civita symbol in 2D.
The space of full symmetric third-order tensors $\mathbb{S}_{(ijk)}$ is $4$-dimensional, meanwhile the space of asymmetric ones $\mathbb{H}^{1}_{\nabla \mathrm{rot}}$ is $2$-dimensional, these spaces are in direct sum.
In the strain-gradient literature \cite{Min64,ME68} the complete symmetric part $ \mathrm{S}_{(ijk)}$, defined:
\ben
S_{(ijk)}=\frac{1}{3}(T_{(ij)k}+T_{(ki)j}+T_{(jk)i})
\een
is related to the \emph{stretch-gradient effects}. Meanwhile the remaining part $V^{\nabla \mathrm{rot}}_{i}$:
\ben
V^{\nabla \mathrm{rot}}_{i}=\epsilon_{ij}\epsilon_{pq}T_{(jp)q}=\left(T_{ipp}-T_{ppi}\right)
\een
is associated to the \emph{the rotation-gradient}. At this stage, this decomposition coincides with the type III formulation of Mindlin  strain-gradient elasticity \cite{Min64,ME68}. According to its words, this  \emph{third form of the theory is the most convenient one for reduction to the theory in which the potential energy-density is a function of the strain and the gradient of the rotation}.  The terms $\mathrm{V}^{\nabla \mathrm{rot}}$ is the components of the  strain-gradient tensor that give rise to couple-stress \cite{Min64}.

In terms of group action, it is important to note that this decomposition\footnote{This decomposition is sometimes known as the Schur decomposition.} is $\mathrm{GL}(2)$-invariant, and that each component is $\mathrm{GL}(2)$-irreducible. In other terms, this decomposition into two \emph{mechanisms} (stretch- and rotation-gradient) is preserved under any invertible transformation. Under $\mathrm{O}(2)$-action each part can further be decomposed into irreducible components by removing their traces:
\begin{itemize}
\item $\mathrm{S}_{(ijk)}$ splits into a septor: $\mathrm{H}^3\in\mathbb{H}^3$ and a vector: $\mathrm{V}^{\nabla \mathrm{str}}_{i}\in\mathbb{H}_{\nabla \mathrm{str}}^{1}$;
\item $\mathrm{V}^{\nabla \mathrm{rot}}_{i}$ can not be split anymore.
\end{itemize}
Hence, the space $\mathbb{S}_{(ijk)}$ is isomorphic to:
\ben
\mathbb{S}_{(ijk)}\cong\mathbb{H}^{3}\oplus\mathbb{H}^{1}_{\nabla \mathrm{str}}
\een
The structure of this decomposition shows that this isomorphism is unique. Doing some algebra we obtain
\ben
S_{(ijk)}=H_{(ijk)}+\frac{1}{4}\left(V^{\nabla \mathrm{str}}_{i}\delta_{(jk)}+V^{\nabla \mathrm{str}}_{j}\delta_{(ik)}+V^{\nabla \mathrm{str}}_{k}\delta_{(ij)}\right)
\een
with 
\ban
H_{(ijk)}&=&S_{(ijk)}-\frac{1}{4}\left(V^{\nabla \mathrm{str}}_{i}\delta_{(jk)}+V^{\nabla \mathrm{str}}_{j}\delta_{(ik)}+V^{\nabla \mathrm{str}}_{k}\delta_{(ij)}\right)\ ;\\
V^{\nabla \mathrm{str}}_{i}&=&S_{(pp)i}=\frac{1}{3}\left(T_{ppi}+2T_{ipp}\right)
\ean
In this formulation $\mathrm{V}^{\nabla \mathrm{str}}$ is the vector part of \emph{the stretch-gradient tensor}. Contrary to the first stage of the decomposition, this new step was not introduced by Mindlin. \\

This decomposition can be summed-up in the following Matryoshka doll fashion:
\bee\label{eq:DecT3}
\mathbb{T}_{(ij)k}=\left(\mathbb{H}^{3}\oplus\mathbb{H}^{1}_{\nabla \mathrm{str}}\right)_{|\mathrm{GL}(2)}\oplus \left(\mathbb{H}^{1}_{\nabla \mathrm{rot}}\right)_{|\mathrm{GL}(2)}
\eee
The parentheses terms are preserved under any invertible transformation, and if this transformation is isometric the harmonic components are furthermore preserved. Hence strain-gradient tensors encode two orthogonal effects: stretch-gradient and rotation-gradient. These effects are canonically defined and preserved in any coordinate system. Once a metric is chosen, the harmonic decomposition corresponds to the Fourier decomposition of these elementary effects. This construction has a meaning for any element of $\mathbb{T}_{(ij)k}$.

\subsubsection*{Second-order tensor approach}
Aside from this first construction, which was based on the algebra of third-order tensors, other constructions can be proposed. The following one is based on the derivation of the harmonic decomposition of a symmetric second-order tensor. As a consequence this construction has a meaning only for tensors constructed in this way.

\ben
\xymatrix{
       &  \mathbb{T}_{(ij)} \ar[ld]_{\mathcal{H}}\ar[rd]^{\mathcal{H}} &  \\
  \mathbb{H}^{2} \ar[d]^{\otimes\V{\nabla}}  &    & \mathbb{H}^{0} \ar[d]^{\otimes\V{\nabla}} \\
  \mathbb{H}^{3}\oplus\mathbb{H}^{1}_{\nabla \mathrm{dev}}&&\mathbb{H}^{1}_{\nabla \mathrm{sph}}}
\een

So the first step is to decompose a second-order symmetric tensor into its deviatoric and spherical parts:
\bee\label{eq:ExpDecSecOrd}
T_{ij}=H_{ij}^{2}+\frac{1}{2}T_{pp}\delta_{ij}\quad \text{with }\ H^{0}=T_{pp}\ \text{and}\  H_{ij}^{2}=T_{ij}-\frac{1}{2}H^{0}\delta_{ij}
\eee
Formally differentiating the harmonic decomposition of $\mathbb{T}_{(ij)}$ we obtain
\ben
\mathbb{H}^{2}\otimes\V{\nabla}=\mathbb{H}^{3}\oplus\mathbb{H}_{\nabla \mathrm{dev}}^{1}\quad;\quad \mathbb{H}^{0}\otimes\V{\nabla}=\mathbb{H}_{\nabla \mathrm{sph}}^{1}
\een
To some extent we have
\ben
\mathbb{T}_{(ij)k}=\mathbb{T}_{(ij)}\otimes\V{\nabla}=\left(\mathbb{H}^{3}\oplus\mathbb{H}_{\nabla \mathrm{dev}}^{1}\right)_{|\mathbb{H}^{2}\otimes\V{\nabla}}\oplus \left(\mathbb{H}_{\nabla \mathrm{sph}}^{1}\right)_{\mathbb{H}^{0}\otimes\V{\nabla}}
\een
But, contrary to the decomposition \eqref{eq:DecT3}, the parentheses terms are not $\mathrm{GL}(2)$-invariant. The first parentheses term is the space of \emph{distortion-gradient} tensors while the other contains \emph{dilatation-gradient} tensors. As $\mathrm{H}^{3}$ is uniquely defined its expression remains unchanged. Therefore attention is focused on the vector parts, doing some algebra we obtain:
\ben
V^{\nabla \mathrm{sph}}_{i}=T_{ppi}\quad ;\quad V^{\nabla \mathrm{dev}}_{i}=\frac{1}{3}\left(2T_{ipp}-T_{ppi}\right)
\een
For $\mathrm{V}^{\nabla \mathrm{sph}}$ the result is direct, for $\mathrm{V}^{\nabla \mathrm{dev}}$ we have:
\ben
\mathrm{Sym}(H_{ij,k}^{2})=S_{ijk}-\frac{1}{6}\left(\delta_{ij}T_{ppk}+\delta_{ki}T_{ppj}+\delta_{jk}T_{ppi}\right)
\een
Therefore, 
\ban
V^{\nabla \mathrm{dev}}_{k}=\frac{1}{3}\left(2T_{kpp}-T_{ppk}\right)
\ean
Those vectors are embedded into the third tensor in the following way:
\ben
T(\mathrm{V}^{\nabla \mathrm{sph}})_{ijk}=\frac{1}{2}V^{\nabla \mathrm{sph}}_{k}\delta_{ij}\quad;\quad T(\mathrm{V}^{\nabla \mathrm{dev}})_{ijk}=\frac{3}{4}\left(V^{\nabla \mathrm{dev}}_{i}\delta_{(jk)}+V^{\nabla \mathrm{dev}}_{j}\delta_{(ik)}-V^{\nabla \mathrm{dev}}_{k}\delta_{(ij)}\right)
\een
It is worth noting that for a Cahn-Hilliard fluid \cite{CH58,Sep89,AIE+13}, i.e. a capillary fluid, the space of strain-gradient tensors reduces to the sole space $\mathbb{H}_{\nabla \mathrm{sph}}^{1}$. At the opposite in (incompressible)  strain-gradient plasticity \cite{SF96,FH01}, only the spaces $\left(\mathbb{H}^{3}\oplus\mathbb{H}_{\nabla\mathrm{dev}}^{1}\right)_{|\mathbb{H}^{2}\otimes\V{\nabla}}$ have to be considered.

\subsection{Synthesis}

If elements of $\mathbb{T}_{(ij)k}$ are considered as gradients of a symmetric second-order tensors, their $\mathrm{O}(2)$-irreducible decompositions can be defined in, at least, two ways\footnote{In the study of the bending gradient plate model, as developed by Lébée and Sab in \cite{LS11a}, an other natural decomposition can be found.}. The first construction is the more general one and is based on the algebra of $\mathbb{T}_{(ij)k}$, meanwhile the second is constructed from the algebra of $\mathbb{T}_{(ij)}$.
Comparing the two decompositions, it appears that the higher-order term $\mathbb{H}^{3}$ is uniquely defined, whereas the vector parts are linear combinations of each others:
\ben
\mathrm{V}^{\nabla \mathrm{str}}=\mathrm{V}^{\nabla \mathrm{dev}}+\frac{2}{3}\mathrm{V}^{\nabla \mathrm{sph}}\quad ;\quad \mathrm{V}^{\nabla \mathrm{rot}}=\frac{3}{2}\mathrm{V}^{\nabla \mathrm{dev}}-\frac{1}{2}\mathrm{V}^{\nabla \mathrm{sph}}
\een
and conversely
\ben
\mathrm{V}^{\nabla \mathrm{sph}}=-\frac{2}{3}\mathrm{V}^{\nabla \mathrm{rot}}+\mathrm{V}^{\nabla \mathrm{str}}\quad ;\quad \mathrm{V}^{\nabla \mathrm{dev}}=\frac{1}{3}(\mathrm{V}^{\nabla \mathrm{str}}+\frac{4}{3}\mathrm{V}^{\nabla \mathrm{rot}})
\een
This analysis gives a mechanical content to the elements of $\mathbb{H}^{3}$:
\begin{itemize}
\item they encode a part of the distortion gradient;
\item they encode a part of the stretch-gradient effect.
\end{itemize}
In the next section it will be demonstrated that $\mathbb{H}^{3}$ is  an eigenspace of $\mathrm{\mathbf{A}}_{\OO(2)}$.
Moreover, the non-uniqueness of the definition of the vector components implies that (stretch- and rotation-gradient) and (distortion- and dilatation-gradient) are coupled phenomena. 

\section{Analysis of the isotropic behavior in harmonic bases}\label{s.MatSys}

Let us get back now to the  matrix representation of the isotropic SGE:
\ben
\mathrm{\mathbf{A}}_{\OO(2)}=
\begin{pmatrix}
 a_{11}& a_{12}& \frac{a_{11}-a_{22}}{\sqrt{2}}-a_{23}& 0                       &0&0\\
      & a_{22}& a_{23}                               &0                 &0     &0\\     
      &      & \frac{a_{11}+a_{22}}{2}-a_{12}       & 0 & 0& 0\\
      &      &                                     & a_{11}& a_{12}& \frac{a_{11}-a_{22}}{\sqrt{2}}-a_{23}\\
      &      &                                     &      & a_{22}& a_{23}\\       
       &&&&&		\frac{a_{11}+a_{22}}{2}-a_{12}
\end{pmatrix}_{\{\mathbf{e}_{1},\mathbf{e}_{2}\}}
\een
This representation is defined in the spatial-based system defined in \S.\ref{ss:OrtOrd}. In this basis, the rotation matrix can be expressed \cite{ABB09}:
\ben
\mathbf{Q}(\theta)=
\begin{pmatrix}
\scriptstyle \mathrm{c}^3(\theta) & \scriptstyle \mathrm{c} (\theta) \mathrm{s}^2(\theta) & \scriptstyle \sqrt{2} \mathrm{c} (\theta) \mathrm{s}^2(\theta) & \scriptstyle -\mathrm{s}^3(\theta) & \scriptstyle -\mathrm{c}^2(\theta) \mathrm{s} (\theta) & \scriptstyle -\sqrt{2} \mathrm{c}^2(\theta) \mathrm{s} (\theta) \\
\scriptstyle \mathrm{c} (\theta) \mathrm{s}^2(\theta) & \scriptstyle \mathrm{c}^3(\theta) & \scriptstyle -\sqrt{2} \mathrm{c} (\theta) \mathrm{s}^2(\theta) & \scriptstyle -\mathrm{c}^2(\theta) \mathrm{s} (\theta) & \scriptstyle -\mathrm{s}^3(\theta) & \scriptstyle \sqrt{2} \mathrm{c}^2(\theta) \mathrm{s} (\theta) \\
\scriptstyle \sqrt{2} \mathrm{c} (\theta) \mathrm{s}^2(\theta) & \scriptstyle -\sqrt{2} \mathrm{c} (\theta) \mathrm{s}^2(\theta) & \scriptstyle \frac{1}{2} (\mathrm{c} (\theta)+\mathrm{c} (3 \theta)) & \scriptstyle -\sqrt{2} \mathrm{c}^2(\theta) \mathrm{s} (\theta) & \scriptstyle \sqrt{2} \mathrm{c}^2(\theta) \mathrm{s} (\theta) & \scriptstyle \mathrm{c} (2 \theta) \mathrm{s} (\theta) \\
\scriptstyle \mathrm{s}^3(\theta) & \scriptstyle \mathrm{c}^2(\theta) \mathrm{s} (\theta) & \scriptstyle \sqrt{2} \mathrm{c}^2(\theta) \mathrm{s} (\theta) & \scriptstyle \mathrm{c}^3(\theta) & \scriptstyle \mathrm{c} (\theta) \mathrm{s}^2(\theta) & \scriptstyle \sqrt{2} \mathrm{c} (\theta) \mathrm{s}^2(\theta) \\
\scriptstyle \mathrm{c}^2(\theta) \mathrm{s} (\theta) & \scriptstyle \mathrm{s}^3(\theta) & \scriptstyle -\sqrt{2} \mathrm{c}^2(\theta) \mathrm{s} (\theta) & \scriptstyle \mathrm{c} (\theta) \mathrm{s}^2(\theta) & \scriptstyle \mathrm{c}^3(\theta) & \scriptstyle -\sqrt{2} \mathrm{c} (\theta) \mathrm{s}^2(\theta) \\
\scriptstyle \sqrt{2} \mathrm{c}^2(\theta) \mathrm{s} (\theta) & \scriptstyle -\sqrt{2} \mathrm{c}^2(\theta) \mathrm{s} (\theta) & \scriptstyle \frac{1}{2} (\mathrm{s} (\theta)-\mathrm{s} (3 \theta)) & \scriptstyle \sqrt{2} \mathrm{c} (\theta) \mathrm{s}^2(\theta) & \scriptstyle -\sqrt{2} \mathrm{c} (\theta) \mathrm{s}^2(\theta) & \scriptstyle \frac{1}{2} (\mathrm{c} (\theta)+\mathrm{c} (3 \theta))
\end{pmatrix}
\een
in which, obviously, $\mathrm{c}$ and $\mathrm{s}$ denotes, respectively, the $\cos$ and $\sin$ function.
The image of a matrix $\mathrm{\mathbf{M}}$ by a rotation is
\ben
\mathrm{\mathbf{M}}^\star=\mathbf{Q}(\theta)\mathrm{\mathbf{M}}\mathbf{Q}^{T}(\theta)
\een
Results of the previous section  allow us to rewrite both the constitutive and rotation matrices in the bases constructed from the harmonic decomposition of $\mathrm{T}_{(ij)k}$. These approaches will finally be compared with the spectral representation of $\mathrm{\mathbf{A}}_{\OO(2)}$.

%Let $\mathrm{End}^S(\mathbb{T}_{(ij)k})$ denotes the space of autoadjoint endomorphism of $\mathbb{T}_{(ij)k}$.
\subsection{First interpretation: stretch and rotation-gradient}

In the first, and more general interpretation of the harmonic decomposition, $\mathbb{T}_{(ij)k}$ is decomposed as follow:
\ben
\mathbb{T}_{(ij)k}=\left(\mathbb{H}^{3}\oplus\mathbb{H}^{1}_{\nabla \mathrm{str}}\right)_{|\mathrm{GL}(2)}\oplus \left(\mathbb{H}^{1}_{\nabla \mathrm{rot}}\right)_{|\mathrm{GL}(2)}
\een
Let us consider the following vectors:
\ben
u_{1}=
\begin{pmatrix}
\frac{1}{2}\\
-\frac{1}{2}\\
-\frac{\sqrt{2}}{2}\\
0\\
0\\
0\\
\end{pmatrix},\ 
u_{2}=
\begin{pmatrix}
0\\
0\\
0\\
\frac{1}{2}\\
-\frac{1}{2}\\
-\frac{\sqrt{2}}{2}\\
\end{pmatrix},\ 
u_{3}=
\begin{pmatrix}
\frac{\sqrt{3}}{2}\\
\frac{\sqrt{3}}{6}\\
\frac{\sqrt{6}}{6}\\
0\\
0\\
0\\
\end{pmatrix} ,\ 
u_{4}=
\begin{pmatrix}
0\\
0\\
0\\
\frac{\sqrt{3}}{2}\\
\frac{\sqrt{3}}{6}\\
\frac{\sqrt{6}}{6}\\
\end{pmatrix} ,\ 
u_{5}=
\begin{pmatrix}
0\\
-\frac{\sqrt{6}}{3}\\
\frac{\sqrt{3}}{3}\\
0\\
0\\
0\\
\end{pmatrix},\ 
u_{6}=
\begin{pmatrix}
0\\
0\\
0\\
0\\
-\frac{\sqrt{6}}{3}\\
\frac{\sqrt{3}}{3}\\
\end{pmatrix}
\een
where $\mathrm{span}(\V{u}_{1},\V{u}_{2})=\mathbb{H}^3$, $\mathrm{span}(\V{u}_{3},\V{u}_{4})=\mathbb{H}^1_{\nabla \mathrm{str}}$  and $\mathrm{span}(\V{u}_{5},\V{u}_{6})=\mathbb{H}^1_{\nabla \mathrm{rot}}$.
The computation of the Gram matrix associated to this family of vectors shows that it constitutes an orthonormal basis for the space $\mathbb{T}_{(ij)k}$.
If $\mathbf{P}_{\mathrm{I}}$ denotes the associate transformation matrix, the operation:
\ben
\mathrm{\mathbf{A}}_{\OO(2)}^{\mathrm{I}}=\mathbf{P}^{T}_{\mathrm{I}}\mathrm{\mathbf{A}}_{\OO(2)}\mathbf{P}_{\mathrm{I}}
\een
expressed the former isotropic operator in the basis of the first harmonic decomposition of $\mathbb{T}_{(ij)k}$. It can be directly observed that $\mathbf{P}_{\mathrm{I}}$ is an element of $\mathrm{O}(6)$ which can not be expressed as an element of $\mathrm{SO}(2)$. In other terms, $\mathbf{P}_{\mathrm{I}}$ is a non-physical orthogonal transformation.
In this new basis, the matrix has the following nice representation:
\bee\label{Eq:BasMecI}
\mathrm{\mathbf{A}}_{\OO(2)}^{\mathrm{I}}=
\begin{pmatrix}
B_{1}&0&0&0&0&0\\
0&B_{1}&0&0&0&0\\
0&0&B_{2}&0&B_{4}&0\\
0&0&0&B_{2}&0&B_{4}\\
0&0&B_{4}&0&B_{3}&0\\
0&0&0&B_{4}&0&B_{3}\\
\end{pmatrix}
\eee
with the four moduli:
\ben
B_{1}=(a_{22}-a_{12} + \sqrt{2} a_{23})\ ;\ B_{2}= \frac{1}{3}(4a_{11}-a_{22}+a_{12} - \sqrt{2} a_{23})\ 
\een
\ben
B_{3}=\frac{1}{6}(a_{11}+5a_{22}-2a_{12}-4\sqrt{2}a_{23})\ ;\quad B_{4}= \frac{\sqrt{2}}{3}(a_{11}-a_{22}-2a_{12} - \sqrt{2} a_{23})
\een
It can directly be observed that the matrix $\mathrm{\mathbf{A}}_{\OO(2)}^{\mathrm{I}}$ is invertible iff:
\ben
B_{1}\left(B_{2}B_{3}-B_{4}^2\right)\neq0
\een
i.e. the modulus $B_{1}$ must be different from 0 but, providing $\left(B_{2}B_{3}-B_{4}^2\right)\neq0$, some other moduli can vanish without making the operator singular. In this new system, the rotation matrix has the following very simple shape:
\bee\label{eq:NewBond}
\mathbf{Q}(\theta)_{\mathrm{I}}=
\begin{pmatrix}
 \mathrm{c} (3 \theta) & \mathrm{s} (3 \theta) & 0 & 0 & 0 & 0 \\
 -\mathrm{s} (3 \theta) & \mathrm{c} (3 \theta) & 0 & 0 & 0 & 0 \\
 0 & 0 & \mathrm{c} (\theta) & -\mathrm{s} (\theta) & 0 & 0 \\
 0 & 0 & \mathrm{s} (\theta) & \mathrm{c} (\theta) & 0 & 0 \\
 0 & 0 & 0 & 0 & \mathrm{c} (\theta) & -\mathrm{s} (\theta) \\
 0 & 0 & 0 & 0 & \mathrm{s} (\theta) & \mathrm{c} (\theta)
\end{pmatrix} 
\eee
which is characteristic of a rotation matrix acting on a space decomposed into harmonic components.

It can directly be observed that, and contrary to classical elasticity (both in 2D and 3D), for the isotropic class, the matrix $\mathrm{\mathbf{A}}_{\OO(2)}$ is not diagonal. Therefore, in the generic situation, a coupling always exists between $\mathbb{H}^{1}_{\nabla \mathrm{str}}$ and $\mathbb{H}^{1}_{\nabla \mathrm{rot}}$. As a consequence, even for isotropic material, pure stretch-gradient or pure rotation-gradient  will both generated double-force and couple-stress \cite{Min64,Ger73}. This representation is similar to the Walpole decomposition as discussed in the literature \cite{Wal84,MB11a,MB13}. Using this representation the mechanical content of the moduli $\{B_{i}\}$ are rather clear:
\begin{itemize}
\item $B_{1}$ and $B_{2}$ are moduli related to the stretch-gradient part of $\eta_{(ij)k}$, $B_{1}$ is linked with the septor part of the stretch-gradient meanwhile $B_{2}$ concerns the vector part;
\item $B_{3}$ is the modulus related to the rotation-gradient part of $\eta_{(ij)k}$;
\item $B_{4}$ is the coupling modulus between the rotation-gradient and the vector part of the stretch gradient.
\end{itemize}
These quantities are interesting for, at least, two reasons. First the computation of the different ratio of these quantities are necessary to quantify the relative importance of the different mechanisms and, at the end, to justify to neglect some of them.
In this view the following quantity, which measures the relative effect of the rotation-gradient to the stretch-gradient can be defined:
\ben
r=\frac{B_{3}}{\sqrt{B_{1}^2+B_{2}^2}}
\een
The second, and closely related, interest is that it makes easy to properly impose kinematic constraints on the behavior. For example, if the material is only sensitive to rotation gradients, the moduli associated with the other mechanisms should be $0$, in other terms 
\ben
B_{1}=B_{2}=B_{4}=0
\een
This condition is satisfy if 
\ben 
a_{11}=a_{12}=0\quad \text{and} \quad a_{22}=-\sqrt{2}a_{23}
\een
In this case
\ben
\mathrm{\mathbf{A}}^{\nabla \mathrm{rot}}_{\OO(2)}=
\begin{pmatrix}
 0& 0& 0& 0                       &0&0\\
      & -\sqrt{2}a_{23}& a_{23}                               &0                 &0     &0\\     
      &      & -\frac{\sqrt{2}}{2}a_{23}       & 0 & 0& 0\\
      &      &                                     & 0& 0& 0\\
      &      &                                     &      &  -\sqrt{2}a_{23}& a_{23}\\       
       &&&&&		-\frac{\sqrt{2}}{2}a_{23} 
\end{pmatrix}
\een
and the behavior is obviously singular. It has to be noted that in \cite{LS11a}, in which the authors studied an higher-order plate model, some very similar degeneracies has been observed. The context is slightly different, since these degeneracies are due to static constraints instead of kinematic ones, but the mechanisms are the same, some moduli equal $0$. In the context of  strain-gradient elasticity a closely related phenomenon has been observed when homogenizing a circular shape \cite{ZMK09,ABB10,MZK12}.

\subsection{Second interpretation: distortion- and dilatation-gradient}

We turn now our interest towards the second harmonic decomposition. In this interpretation we have:
\ben
\mathbb{T}_{(ij)k}=\mathbb{T}_{(ij)}\otimes\V{\nabla}=\left(\mathbb{H}^{3}\oplus\mathbb{H}_{\nabla \mathrm{dev}}^{1}\right)_{|\mathbb{H}^{2}\otimes\V{\nabla}}\oplus \left(\mathbb{H}_{\nabla \mathrm{sph}}^{1}\right)_{\mathbb{H}^{0}\otimes\V{\nabla}}
\een
The mechanical information contained in $\mathbb{H}^3\oplus\mathbb{H}^1_{\nabla \mathrm{dev}}$ is related to the gradient of the deviator, meanwhile $\mathbb{H}^1_{\nabla \mathrm{sph}}$ is the gradient of the spherical part of a second order tensor. Let us consider the following vectors
\ben
v_{1}=
\begin{pmatrix}
\frac{1}{2}\\
-\frac{1}{2}\\
-\frac{\sqrt{2}}{2}\\
0\\
0\\
0\\
\end{pmatrix}\ ,\ 
v_{2}=
\begin{pmatrix}
0\\
0\\
0\\
\frac{1}{2}\\
-\frac{1}{2}\\
-\frac{\sqrt{2}}{2}\\
\end{pmatrix}\ ,\ 
v_{3}=
\begin{pmatrix}
\frac{1}{2}\\
-\frac{1}{2}\\
\frac{\sqrt{2}}{2}\\
0\\
0\\
0\\
\end{pmatrix}\ ,\ 
v_{4}=
\begin{pmatrix}
0\\
0\\
0\\
\frac{1}{2}\\
-\frac{1}{2}\\
\frac{\sqrt{2}}{2}\\
\end{pmatrix}\ ,\ 
v_{5}=
\begin{pmatrix}
\frac{\sqrt{2}}{2}\\
\frac{\sqrt{2}}{2}\\
0\\
0\\
0\\
0\\
\end{pmatrix}\ ,\ 
v_{6}=
\begin{pmatrix}
0\\
0\\
0\\
\frac{\sqrt{2}}{2}\\
\frac{\sqrt{2}}{2}\\
0\\
\end{pmatrix}
\een
where $\mathrm{span}(\V{v}_{1},\V{v}_{2})=\mathbb{H}^3$, $\mathrm{span}(\V{v}_{3},\V{v}_{4})=\mathbb{H}^1_{\nabla \mathrm{dev}}$  and $\mathrm{span}(\V{v}_{5},\V{v}_{6})=\mathbb{H}^1_{\nabla \mathrm{sph}}$.
The computation of the Gram matrix associated to this family of vectors shows that it constitutes an orthonormal basis for the space $\mathbb{T}_{(ij)k}$. If we note $\mathbf{P}_{\mathrm{II}}$ the associate transformation matrix, the operation:
\ben
\mathrm{\mathbf{A}}_{\OO(2)}^{\mathrm{II}}=\mathbf{P}^{T}_{\mathrm{II}}\mathrm{\mathbf{A}}_{\OO(2)}\mathbf{P}_{\mathrm{II}}
\een
expresses the former isotropic operator in the basis of the second harmonic decomposition of $\mathbb{T}_{(ij)k}$.
In this new basis, the matrix has the following nice representation:
%In the most general situation this base is constructed by
%\ben
%b_{ij}=\frac{\delta_{ij}+\sqrt{2}(1-\delta_{ij})}{2}(v_{i}\otimes v_{j}+v_{j}\otimes v_{i})
%\een
%but for tensor belonging to high symmetry classes, only few of this vectors are necessary. Let consider the matrix \eqref{eq:D6mat}, using the change of variable defined by $\{(v_{i})_{j}\}$ in the new system it takes the following form:
\bee\label{Eq:BasMecII}
\mathrm{\mathbf{A}}_{\OO(2)}^{\mathrm{II}}=
\begin{pmatrix}
C_{1}&0&0&0&0&0\\
0&C_{1}&0&0&0&0\\
0&0&C_{2}&0&C_{4}&0\\
0&0&0&C_{2}&0&C_{4}\\
0&0&C_{4}&0&C_{3}&0\\
0&0&0&C_{4}&0&C_{3}\\
\end{pmatrix}
\eee
with
\ben
C_{1}=(a_{22} - a_{12} + \sqrt{2} a_{23})=B_{1} \quad;\quad C_{2}= (a_{11} - a_{12} - \sqrt{2} a_{23})
\een
\ben
C_{3}= \frac{1}{2}(a_{11}+a_{22}+2a_{12})\ ;\ C_{4}= \frac{\sqrt{2}}{2}(a_{11} - a_{22})\ 
\een
As in the previous case, the matrix is invertible provided:
\ben
C_{1}\left(C_{2}C_{3}-C_{4}^2\right)\neq0
\een
And, in this basis, the rotation matrix has the same simple shape as \eqref{eq:NewBond}. Using this representation the mechanical content of the moduli $\{C_{i}\}$ are rather clear:
\begin{itemize}
\item $C_{1}$ and $C_{2}$ are moduli related to the distortion-gradient part of $\eta_{(ij)k}$, $C_{1}$ is linked with the septor part of the distortion-gradient meanwhile $C_{2}$ concerns the vector part;
\item $C_{3}$ is the modulus related to the dilatation-gradient part of $\eta_{(ij)k}$;
\item $C_{4}$ is the coupling modulus between the dilatation-gradient and the vector part of the distortion-gradient. Furthermore it gives a direct interpretation to the matrix coefficient $a_{22}$ since the coupling is  as more important as $a_{22}$ is different from $a_{11}$.
\end{itemize}
In the case of a material only sensitive to the dilatation gradient (a solid equivalent to a Cahn-Hilliard fluid \cite{Sep89,AIE+13}), the moduli associated with the other mechanisms should be $0$, in other terms 
\ben
C_{1}=C_{2}=C_{4}=0
\een
This condition is satisfy if 
\ben 
a_{11}=a_{12}=a_{22}\quad \text{and} \quad a_{23}=0
\een
In this case the constitutive matrix reduces to
\ben
\mathrm{\mathbf{A}}^{\nabla \mathrm{sph}}_{\OO(2)}=
\begin{pmatrix}
 a_{11}& a_{11}& 0& 0                       &0&0\\
      & a_{11}& 0                               &0                 &0     &0\\     
      &      & 0      & 0 & 0& 0\\
      &      &                                     & a_{11}& a_{11}& 0\\
      &      &                                     &      & a_{11}& 0\\       
       &&&&&	0
\end{pmatrix}
\een

\subsection{Comparison with the spectral decomposition}

In this last subsection the former \emph{Walpole-type} decompositions are compared with the spectral decomposition of $\mathrm{\mathbf{A}}_{\OO(2)}$ \cite{Kel56,Fra11}. The spectral decomposition is an important decomposition since the eigenvalues both determine whether the operator is singular or not and, if so, the condition number of the linear application. As the characteristic polynomial of a block diagonal matrix is the product of the characteristic polynomial of each block, it is obvious, looking at the matrix \eqref{Aiso}, that, in generic situations,  $\mathrm{\mathbf{A}}_{\OO(2)}$ has three different eigenvalues each of them with multiplicity $2$.
A direct computation gives the expression of the associated diagonal matrix:
\bee\label{Eq:BasSpe}
\mathrm{\mathbf{A}}_{\OO(2)}^{\mathrm{Sp}}=
\begin{pmatrix}
\lambda_{1}&0&0&0&0&0\\
0&\lambda_{1}&0&0&0&0\\
0&0&\lambda_{2}&0&0&0\\
0&0&0&\lambda_{2}&0&0\\
0&0&0&0&\lambda_{3}&0\\
0&0&0&0&0&\lambda_{3}\\
\end{pmatrix}
\eee
with
\ben
\lambda_{1}=B_{1}=C_{1} \quad;\quad \lambda_{2}= T+\sqrt{\Delta}\ ;\ \lambda_{3}= T-\sqrt{\Delta}
\een
The two remaining eigenvalues can be defined algebraically in terms of $B_{2}$, $B_{3}$ and $B_{4}$ (conv. $C_{2}$, $C_{3}$ and $C_{4}$).
\ban
2T&=&B_{2}+B_{3}=C_{2}+C_{3}\\
\Delta&=&(B_{2}+B_{3})^2-4(B_{2}B_{3}-B_{4}^2)=(C_{2}+C_{3})^2-4(C_{2}C_{3}-C_{4}^2)
\ean
Denoting $\mathbb{E}_{\lambda_{i}}$ the eigenspace associated to $\lambda_{i}$, the space $\mathbb{T}_{(ij)k}$ is divided into 3 orthogonal bi-dimensional eigenspaces.
\bee\label{eq:EigO2}
\mathbb{T}_{(ij)k}=\mathbb{E}_{\lambda_{1}}\oplus\mathbb{E}_{\lambda_{2}}\oplus\mathbb{E}_{\lambda_{3}},\quad \dim\mathbb{E}_{\lambda_{i}}=2
\eee
A direct analysis reveals that, indeed, $\mathbb{E}_{\lambda_{1}}=\mathbb{H}^{3}$. In the basis constituted of the eigenvectors, the rotation matrix is block diagonal and have the same expression as before (c.f. equation \eqref{eq:NewBond}), as a consequence both $\mathbb{E}_{\lambda_{2}}$ and $\mathbb{E}_{\lambda_{3}}$ are associated with some $\mathbb{H}^{1}$.

\section{Conclusion}
A general construction process of second order isotropic moduli has been proposed. As a result it appears that many sets of them can be defined, each of them constituted of 4 moduli: 3 associated with 2 distinct mechanisms and the last one coupling these mechanisms. Hence, it has been shown that, and contrary to classical elasticity, even for the isotropy class second order elasticity is a coupled behavior\footnote{As demonstrated in \cite{Auf13b} this fact indeed occurs each time the number of isotropic coefficients exceeds the number of elementary spaces in the harmonic decomposition of the strain-gradient tensor.}. 
These results have been supplied by two explicit constructions these moduli. The interest of ours results are two-fold:
\begin{itemize}
\item A mechanical content has been given to higher-order moduli. Their knowledge are important both to analyze the second-order kinematic and to impose kinematic constrains;
\item Furthermore this construction is general and can be applied in many other situations. For example, the analysis can easily be extended to micromorphic continua \cite{}. This point will be considered in a forthcoming work.
\end{itemize}
We hope that these moduli (and the construction process) will be useful for forthcoming investigations on generalized continuum mechanics.

\appendix
\section*{Appendix}

In this appendix explicit decompositions of $\mathrm{T}_{(ij)k}$ are provided. We choose the following matrix representation:
\ben
\mathrm{\mathbf{T}}=
\begin{pmatrix}
T_{111}&T_{112}\\
T_{221}&T_{222}\\
\sqrt{2}T_{122}&\sqrt{2}T_{121}\\
\end{pmatrix}
\een

\subsection*{Affine decomposition ($\mathrm{GL}(2)$-invariant)}

\noindent$\bullet$\emph{The stretch-gradient tensor:}\\ 
\ben
\mathrm{\mathbf{S}}=
\begin{pmatrix}
S_{1}&S_{3}\\
S_{4}&S_{2}\\
\sqrt{2}S_{4}&\sqrt{2}S_{3}\\
\end{pmatrix}
=
\begin{pmatrix}
T_{111}&\frac{1}{3}(T_{112}+2T_{121})\\
\frac{1}{3}(T_{221}+2T_{122})&T_{222}\\
\frac{\sqrt{2}}{3}(T_{221}+2T_{122})&\frac{\sqrt{2}}{3}(T_{112}+2T_{121})\\
\end{pmatrix}
\een

\noindent$\bullet$\emph{The rotation-gradient tensor:}\\ 

\noindent As a vector:
\ben
\mathrm{\mathbf{V}}^{\nabla \mathrm{rot}}=
\begin{pmatrix}
T_{122}-T_{221}\\
T_{121}-T_{112}\\
\end{pmatrix}\een
and embedded into $\mathrm{\mathbf{R}}$:
\ben
\mathrm{\mathbf{R}}=
\begin{pmatrix}
0&-\frac{2}{3}V^{\nabla \mathrm{rot}}_{2}\\
-\frac{2}{3}V^{\nabla \mathrm{rot}}_{1}&0\\
\frac{\sqrt{2}}{3}V^{\nabla \mathrm{rot}}_{1}&\frac{\sqrt{2}}{3}V^{\nabla \mathrm{rot}}_{2}\\
\end{pmatrix}
=
\begin{pmatrix}
0&-\frac{2}{3}(T_{121}-T_{112})\\
-\frac{2}{3}(T_{122}-T_{221})&0\\
\frac{\sqrt{2}}{3}(T_{122}-T_{221})&\frac{\sqrt{2}}{3}(T_{121}-T_{112})\\
\end{pmatrix}
\een

\subsection*{Harmonic decomposition ($\mathrm{O}(2)$-invariant)}

\noindent$\bullet$\emph{Vector part of the stretch-gradient tensors:}\\

\noindent As a vector:
\ben
\mathrm{\mathbf{V}}^{\nabla \mathrm{str}}=
\begin{pmatrix}
T_{111}+\frac{1}{3}(T_{221}+2T_{122})\\
T_{222}+\frac{1}{3}(T_{112}+2T_{121})\\
\end{pmatrix}
\een
and embedded into $\mathrm{\mathbf{T}}$:
\ben
\mathrm{\mathbf{T}}(\mathrm{\mathbf{V}}^{\nabla \mathrm{str}})=
\begin{pmatrix}
\frac{3}{4}V^{\nabla \mathrm{str}}_{1}&\frac{1}{4}V^{\nabla \mathrm{str}}_{2}\\
\frac{1}{4}V^{\nabla \mathrm{str}}_{1}&\frac{3}{4}V^{\nabla \mathrm{str}}_{2}\\
\frac{\sqrt{2}}{4}V^{\nabla \mathrm{str}}_{1}&\frac{\sqrt{2}}{4}V^{\nabla \mathrm{str}}_{2}\\
\end{pmatrix}
\een

\noindent$\bullet$\emph{$3^{rd}$-order deviator of any  strain-gradient tensor:}\\

\noindent As a vector:
\ban
\mathrm{\mathbf{{H}}}^{3}=\begin{cases}
\frac{1}{4}(T_{111}-T_{221}-2T_{122})\\
\frac{1}{4}(T_{222}-T_{112}-2T_{121})\\
\end{cases}
\ean
and embedded into $\mathrm{\mathbf{T}}$:
\ben
\mathrm{\mathbf{T}}(\mathrm{\mathbf{{H}}}^{3})=
\begin{pmatrix}
H_{1}&-H_{2}\\
-H_{1}&H_{2}\\
-\sqrt{2}H_{1}&\sqrt{2}H_{2}\\
\end{pmatrix}=
\begin{pmatrix}
\frac{1}{4}(T_{111}-T_{221}-2T_{122})&-\frac{1}{4}(T_{222}-T_{112}-2T_{121})\\
-\frac{1}{4}(T_{111}-T_{221}-2T_{122})&\frac{1}{4}(T_{222}-T_{112}-2T_{121})\\
\frac{-\sqrt{2}}{4}(T_{111}-T_{221}-2T_{122})&\frac{\sqrt{2}}{4}(T_{222}-T_{112}-2T_{121})\\
\end{pmatrix}\\
\een

%\subsubsection{$V^{\nabla rot}$}
%%\ben
%%R_{ij}=
%%\begin{pmatrix}
%%T_{123}-T_{132}&T_{223}-T_{232}&T_{233}-T_{332}\\
%%T_{131}-T_{113}&T_{231}-T_{123}&T_{331}-T_{313}\\
%%T_{112}-T_{121}&T_{122}-T_{221}&T_{132}-T_{231}
%%\end{pmatrix}
%%\een
%%
%%\ban
%%R_{1}=R_{(122)}&=&\frac{1}{3}R_{32}=\frac{1}{3}\left(T_{122}-T_{221}\right)\\
%%R_{2}=R_{(133)}&=&-\frac{1}{3}R_{23}=-\frac{1}{3}\left(T_{331}-T_{313}\right)\\
%%R_{3}=R_{(121)}&=&-\frac{1}{3}R_{31}=-\frac{1}{3}\left(T_{112}-T_{121}\right)\\
%%R_{4}=R_{(233)}&=&\frac{1}{3}R_{13}=\frac{1}{3}\left(T_{233}-T_{332}\right)\\
%%R_{5}=R_{(131)}&=&-\frac{1}{3}R_{21}=\frac{1}{3}\left(T_{131}-T_{113}\right)\\
%%R_{6}=R_{(232)}&=&\frac{1}{3}R_{12}=-\frac{1}{3}\left(T_{223}-T_{232}\right)\\
%%R_{7}=R_{(123)}&=&\frac{1}{3}\left(R_{11}-R_{22}\right)=\frac{1}{3}\left(2T_{123}-T_{132}-T_{231}\right)\\
%%R_{8}=R_{(231)}&=&\frac{1}{3}\left(R_{22}-R_{33}\right)=\frac{1}{3}\left(2T_{231}-T_{132}-T_{123}\right)\\
%%\ean
%
%\ben
%V^{\nabla rot}
%\begin{pmatrix}
%T_{122}-T_{221}\\
%T_{121}-T_{112}\\
%\end{pmatrix}
%\een
%
%\ban
%[\tTd{T}(V^{\nabla rot})]&=&
%\begin{pmatrix}
%0&-\frac{2}{3}V^{\nabla rot}_{2}\\
%-\frac{2}{3}V^{\nabla rot}_{1}&0\\
%\frac{\sqrt{2}}{3}V^{\nabla rot}_{1}&\frac{\sqrt{2}}{3}V^{\nabla rot}_{2}\\
%\end{pmatrix}\\
%&=&
%\begin{pmatrix}
%0&-\frac{2}{3}(T_{121}-T_{112})\\
%-\frac{2}{3}(T_{122}-T_{221})&0\\
%\frac{\sqrt{2}}{3}(T_{122}-T_{221})&\frac{\sqrt{2}}{3}(T_{121}-T_{112})\\
%\end{pmatrix}
%\ean

\noindent$\bullet$\emph{Dilatation-gradient vector:}\\

\noindent As a vector:
\ben
\mathrm{\mathbf{V}}^{\nabla \mathrm{sph}}=
\begin{pmatrix}
T_{111}+T_{221}\\
T_{222}+T_{112}\\
\end{pmatrix}
\een
and embedded in $\mathrm{\mathbf{T}}$:
\ben
\mathrm{\mathbf{T}}(\mathrm{\mathbf{V}}^{\nabla \mathrm{sph}})=
\begin{pmatrix}
\frac{1}{2}V^{\nabla \mathrm{sph}}_{1}&\frac{1}{2}V^{\nabla \mathrm{sph}}_{2}\\
\frac{1}{2}V^{\nabla \mathrm{sph}}_{1}&\frac{1}{2}V^{\nabla \mathrm{sph}}_{2}\\
0&0\\
\end{pmatrix}=
\begin{pmatrix}
\frac{1}{2}(T_{111}+T_{221})&\frac{1}{2}(T_{222}+T_{112})\\
\frac{1}{2}(T_{111}+T_{221})&\frac{1}{2}(T_{222}+T_{112})\\
0&0\\
\end{pmatrix}
\een

\noindent$\bullet$\emph{Distortion-gradient vector:}\\

\noindent As a vector:
\ben
\mathrm{\mathbf{V}}^{\nabla \mathrm{dev}}=
\begin{pmatrix}
\frac{1}{3}(T_{111}-T_{221}+2T_{122})\\
\frac{1}{3}(T_{222}-T_{112}+2T_{121})\\
\end{pmatrix}
\een
and embedded in $\mathrm{\mathbf{T}}$:
\ben
\mathrm{\mathbf{T}}(\mathrm{\mathbf{V}}^{\nabla \mathrm{dev}})
\begin{pmatrix}
\frac{3}{4}V^{\nabla \mathrm{dev}}_{1}&-\frac{3}{4}V^{\nabla \mathrm{dev}}_{2}\\
-\frac{3}{4}V^{\nabla \mathrm{dev}}_{1}&\frac{3}{4}V^{\nabla \mathrm{dev}}_{2}\\
\frac{3\sqrt{2}}{4}V^{\nabla \mathrm{dev}}_{1}&\frac{3\sqrt{2}}{4}V^{\nabla \mathrm{dev}}_{2}\\
\end{pmatrix}\een

\bibliographystyle{spmpsci}      % mathematics and physical sciences
\bibliography{elasticity2013}   % name your BibTeX data base

%% Non-BibTeX users please use
%\begin{thebibliography}{}
%%
%% and use \bibitem to create references. Consult the Instructions
%% for authors for reference list style.
%\end{thebibliography}

\end{document}